\documentclass[useAMS,usenatbib]{mn2e}
\usepackage{graphicx}

\title[An analysis of the distribution of background star
   polarization in dark clouds]{An analysis of the distribution of background star
   polarization in dark clouds}
\author[A. K. Sen, T. Mukai, R. Gupta and H. S. Das]{A. K. Sen$^{1}$\thanks{E-mail:
asokesen@sancharnet.in }, T. Mukai $^{2}$\thanks {E-mail:
mukai@kobe-u.ac.jp } , R. Gupta $^3$\thanks{E-mail:
rag@iucaa.ernet.in} and
H. S. Das$^{1}$\\
$^{1}$Department of Physics, Assam University,
              Silchar 788011, Assam, India\\
$^{2}$Dept. of Earth and Planetary Sciences, Faculty of Science,
          Kobe University, Nada 657, Kobe,
         Japan\\
$^{3}$IUCAA, Post Bag 4, Ganeshkhinde, Pune 411007, India\\
}
\begin{document}

\date{Accepted ~~~~~~~~~ Received ~~~~~~~~~~~~~~ in original form }

\pagerange{\pageref{firstpage}--\pageref{lastpage}} \pubyear{2002}

\maketitle

\label{firstpage}

\begin{abstract}
The polarization observed for  stars background to dark clouds
(Bok Globules)  is often used as diagnostic to study the ongoing
star formation processes in these clouds. Such polarization maps
in the optical have been reported for eight nearby clouds CB3,
CB25, CB39, CB52, CB54, CB58, CB62 and CB246 in one of our
previous work (Sen et al 2000). With a  view to understand the
origin of this polarization, in the present work attempts are made
to look for any possible relation between this observed
polarization and other physical parameters in the cloud (like
temperature, turbulence etc.).  The observed polarization  does
not seem to be clearly related to the dust and gas temperatures
($T_d$ and $T_g$) in the cloud  as expected from Davis-Greenstein
grain alignment mechanism (Davis \& Greenstein 1952). However, the
average observed polarization ($p_{av}$) appears to be related to
the turbulence $\Delta V$ (measured by $^{12}CO$ line width) by
the mathematical relation $p_{av}=2.95~~ exp(-0.24 \Delta V)$. The
possible relation between the direction of polarization vector and
other physical parameters are also discussed. For this analysis in
addition to the data on above eight dark clouds, the data on CB4
(Kane et al. 1995) are also included for comparison.

In order to study the spatial distribution of the degree of
polarization and position angles across the different parts of the
cloud  a simple model is proposed,  where the cloud has been
assumed to be a simple dichroic polarizing sphere and  the light
from the background star first passes through the IS medium and
then through the cloud, before reaching the observer. One finds
this simple model can  explain to a reasonable extent the observed
spatial (radial) dependence of the value of $p$ for two of the
clouds (CB25 , CB39), but for rest of the clouds the model fails.
However, through this model one can explain why the polarization
($p$) need not always increase with total extinction $A_v$ as one
moves in the deeper interior part of the cloud.
\end{abstract}


\section{Introduction}
The small compact dark clouds or 'Bok Globules' as they are also
known as, are believed to be the ideal sites for star formation
(Bok \& Reilly 1947 ). Such clouds have been catalogued by Bernard
(1927), Lynds (1962) and more recently by Clemens \& Barvainis
(1988).

These clouds are undergoing gravitational collapse and eventually
may form stars. The ambient magnetic field plays a key role in the
collapse dynamics by directing the outflows, impeding the plasma
movement across magnetic field and in many other ways. Owing to
this, there have been several attempts in past to measure strength
and geometry of the magnetic field within the cloud. Astronomers
have been using background star polarimetry as a tool to
understand the ambient magnetic field and study the star formation
dynamics in the cloud (Vrba et al 1981; Joshi et al. 1985; Goodman
et al 1989; Myers \& Goodman 1991; Kane et al 1995; Sen et al.
2000, to mention a few). This technique has an underlying
assumption that, the light from the background stars are scattered
in the forward  direction by the magnetically aligned  dichroic
dust grains in the cloud. Davis \& Greenstein (1952) first worked
out a  procedure showing how grain alignments are possible by
magnetic field. Several modifications of this mechanism and
various other alignment mechanisms are presently discussed in the
literature (for a detail review on this please see Lazarian et al.
(1997)).

It is normally expected that, grains which cause polarization,
should also be responsible for the extinction observed for the
background stars. However, Goodman et al (1995) observed a lack of
dependence of polarization with extinction  and this has
questioned the validity of polarization as a tracer of magnetic
field in these clouds. More recently Sen et al (2000) have mapped
eight star forming clouds CB3, CB25, CB39 , CB 52, CB54, CB58,
CB62 and CB246 in white light polarization and commented on  the
possible star formation dynamics there.

With the above background, in this paper a detail analysis of the
polarization images of the above eight clouds is carried out.
Attempts were made to understand whether the ambient physical
parameters like temperature and turbulence have any role on the
observed polarization value. Further the projected angular
distance (henceforth 'radial distance') of the background stars
from the cloud center were estimated, for the eight clouds as
observed by Sen et al (2000). Hence the data was analysed, to find
whether the polarization values observed for these stars are
anyway related with these distances ?

\section{The statical distribution of the degree of polarization
and position angle in a given cloud} It is well known that the
observed polarization  is a positive definite quantity and instead
of Gaussian distribution it follows Ricean distribution given by
(Simon \& Stewart 1984)

\begin{equation}
F(p,p_0)= {p \over {\sigma_p}} I_0 (\frac{p p_0}{\sigma_p^2})exp
^{(\frac {p^2+p_0^2}{2 \sigma _p^2})}
\end{equation}

where $p_0$  is the true value of fractional polarization being
estimated by $p$ and $I_0$ is the modified Bessel function of
order zero. There are several schemes available  for de-biasing
these data, however none of these schemes are fully satisfactory.
There is a Rice factor $[1-(\sigma_p^2/p^2)]^{(1/2)}$ which is
often used to de-bias such data. By multiplying each observed
polarization value by the Rice factor,  the polarization values
are corrected for their non-Gaussin nature. In Fig 1  histogram
plots showing number of stars within a given range of Rice
corrected polarization values for each cloud are made. The
position angle or direction vector of observed polarization
($\theta$) values of all stars can also be considered for a
similar analysis. It is normally assumed that the polarization is
caused due to ambient magnetic field in the cloud with the
direction of polarization lying along the direction of the
magnetic field  in the cloud. The direction of ambient magnetic
field in a given cloud can be assumed as the direction of the
projection of galactic plane in that part of the cloud (denoted by
$\theta_G$). Now  in order to study the distribution of observed
$\theta $ values in different clouds, in Fig. 2 similar histogram
plots are made showing number of stars observed in a given range
of $\theta $ values.

As can be seen from Fig. 1, the clouds CB3, CB52, CB58 and CB246
show a tendency for  bimodal polarization (Rice corrected)
distributions. For other clouds  only one peak in the number
distribution is observed. The bimodal distribution can be
explained in a number of ways. As discussed in detail by Vrba et
al. (1988) and also commented by Myers \& Goodman (1995) (for a
similar study on CB4), the low polarization component may be
arising out of the foreground stars and the high polarization
component may be due to the background stars. However,  in a dark
cloud with a given sample of stars, some stars are neither
foreground nor background to the cloud, but lying a bit outside
the periphery of the cloud. This happens because shape of the
cloud is mostly irregular and does not evenly cover the area of
the rectangular detector. So these stars also contribute to the
polarization data, and represent simple interstellar polarization.
Such stars probably contribute largely to the second Gaussian
component, as in the present case all the clouds are quite nearby
and one should not have many stars foreground to the clouds. It is
also likely that (i)the polarization produced within the cloud has
direction different from  that produced in the interstellar
medium. In the IS medium  one should have polarization mostly
aligned along the direction of galactic magnetic field (coinciding
with the direction of galactic plane ) or (ii) within the same
cloud itself there may be no-uniform magnetic fields. Such
features can be studied from the histogram plot of $ \theta $.
Myers \& Goodman (1995) made a very detailed analysis on the
dispersion in the direction of polarization for 15 dark clouds,
five clusters and six complexes. It was shown that the bimodal
distributions can be explained, through a model, where there exist
two components in magnetic field one uniform and another
non-uniform. The non-uniform part has an isotropic probability
distribution of direction, a Gaussian distribution of amplitudes
and N correlation lengths along the line of sight. This model was
applied to the cloud L1755 by Goodman et al. (1995) to explain the
distribution of direction of polarization vectors.

As can be seen from Fig 2., almost all the clouds show a single
peak in the distribution of $\theta$ values, which is somewhat
very close to the direction of galactic magnetic field
($\theta_G$). In clouds CB52 and CB58 there may be small exception
showing two peaks in $\theta$ distribution, but it is not very
significant. A closer look at the histogram depicts that the peak
in  $\theta$ values for all the clouds lies within $1 \sigma
_\theta$ (as listed in Table 1) from the direction of galactic
plane $\theta _G$ (representing magnetic field). A similar
conclusion can also be arrived at by looking at Table 1, where one
finds the $ \sigma _\theta$  and $|\theta _G - \theta |$ values
are very close to each other. This observation suggests that the
direction of IS magnetic field and that of the magnetic field
within the cloud (responsible for grain alignment) may be the same
or these two differ only within $1 \sigma$. The field responsible
for the alignment of grains, therefore, can be assumed to be
related to the galactic magnetic field.

\begin{figure}
\includegraphics[width=8cm] {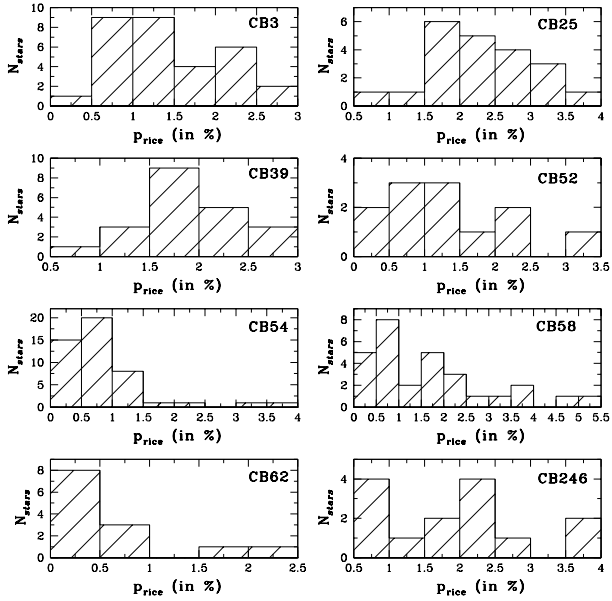}
 \caption{Histogram showing the number
($N_{stars}$) distribution of stars having Rice corrected
polarization  ($p_{rice}$) values in different ranges for various
clouds.}
 \label{FigVibStab}
\end{figure}

\begin{figure}
   \centering
\includegraphics[width=8cm]{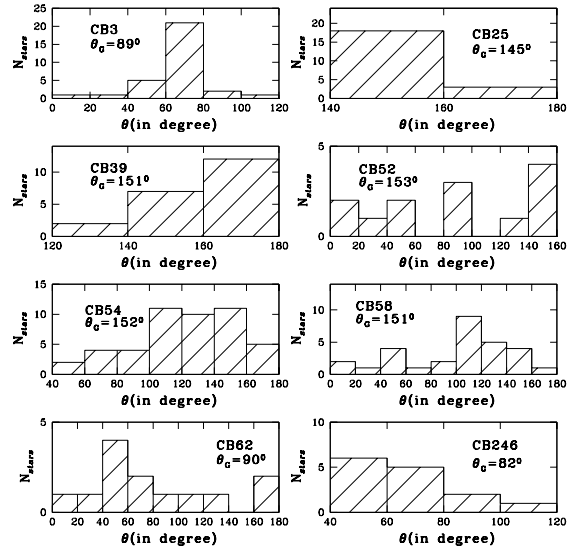}
      \caption{Histogram showing the number ($N_{stars}$) distribution of stars having position
      angle ($\theta$) values in different ranges for various clouds
}
         \label{FigVibStab}
   \end{figure}

\section[]{Observed polarization and ambient physical conditions in the cloud }

\subsection{The dependence of observed polarization on dust and gas temperature }

The light from stars background to the cloud is generally found to
be polarized. This happens due to the scattering of the light from
the background stars by the aligned dichroic grains present in the
cloud. It is believed that the alignment is resulted from an
interaction between the rotational dynamics of the grains and the
ambient magnetic field. This mechanism called paramagnetic
relaxation was originally suggested by Davis \& Greenstein (1951).
It can be shown that the percentage of polarization ($p \%$) as
expected by this mechanism can be expressed as (Vrba et al. 1981):

\begin{equation}
p (\%)= 67 F A_v
\end{equation}

where $A_v$ is total visual extinction and the expression for $F$
can be found from Jones \& Spitzer (1967) :

\begin{equation}
F=\frac{\chi ^{''} B^2 }{75 a \omega n}(\frac{2 \pi}{m k
T_g})^{1/2} (\gamma -1) (1-T_d/ T_g)
\end{equation}

where $\chi ^{''}$ is the imaginary part of the complex
susceptibility of the grains, $\omega$ is the angular velocity of
rotation, $T_d$ and $T_g$ are the dust and gas (kinetic)
temperatures, $B$ is magnetic field,  n is gas density in the
vicinity of grain (generally taken as Hydrogen gas density), $m$
is gas molecular mass, k is Boltzman constant, $\gamma
=(1/2)[(b/a)^2+1]$, $b$ and $a$ are short and  long axes of the
grains . Further it is known that (Davis \& Greenstein 1951;
Purcell 1979) :

$$\frac{\chi ^{''}}{\omega}= 2.6 ~10^{-12}T_d ^{-1}$$

Therefore one can write a simplified expression for $p(\%)$ as :

\begin{equation}
p(\%)\sim \frac{ B^2 }{n}
\frac{1}{\sqrt{T_g}}(\frac{1}{T_d}-\frac{1}{T_g})A_v
\end{equation}

The total extinction   $A_v$ in a cloud can be related to the gas
(hydrogen) density in the cloud. The relation $A_v \sim \sigma_H$,
with $\sigma _H$ as gas coloumn density, seems to be true for all
parts of the cloud and this relation has been experimentally
verified except at very high opacities (Jenkins \& Savage 1974).
Subsequently many authors (Dickman 1978; Gerakins et al. 1995)
used such a relation to study various physical parameters of
clouds. Thus following Vrba et al.(1981) one may write $n (atoms~~
cm^{-3}) = 532 * A_v/ l(~~ pc) $, where $l$ is the path length
from the background star to the observer. Therefore assuming
classical Davis \& Greenstein  mechanism one may obtain from the
Eqn. (4):

\begin{equation}
p(\%) \sim B^2  \frac{1}{\sqrt{T_g}}(\frac{1}{T_d}-\frac{1}{T_g})l
\end{equation}

The background stars have been randomly selected in the present
sample. Therefore, one may assume   the path lengths from the
various background stars to the clouds to be of same order.
Further these clouds are very near, so one may also assume that
the path lengths (between the cloud and the observer) will be very
small and will vary little from cloud to cloud. Under these
circumstances one may write:

$$
p(\%) \sim constant +
\frac{B^2}{\sqrt{T_g}}(\frac{1}{T_d}-\frac{1}{T_g}) b
$$

where the  '$constant$' term above is equivalent to the
contribution in polarization of the path length in the region
outside the cloud. This contribution has been assumed to be a
constant, as these path lengths are of same order and the
variations in temperatures of the interstellar medium outside the
various clouds will be very small. The other quantity $b$ is the
path length through the cloud and assuming the cloud to be
spherical in shape, it can be expressed by the relation $b= 2
\sqrt {R_0^2 -r^2}$, where $R_0$ is the projected angular radius
of the cloud and $r$ is the projected angular distance of the
background star from the cloud centre (details of this geometry
are also discussed in section 4.3).  To find the average of
observed polarization in a cloud one has to  consider the average
value of $b$ in a cloud, and it can be easily shown than $b_{av}
\sim R_0$. Hence one will write

$$
p_{av}(\%) \sim constant +
\frac{B^2}{\sqrt{T_g}}(\frac{1}{T_d}-\frac{1}{T_g})R_0
$$

The above equation has been obtained with several assumptions and
may represent a highly  simplistic and ideal situation. Also  the
above equation can not be simplified any further. However, in
order to explore a possible dependence of observed polarization on
temperature, as a very crude approximation one may further assume
that, the clouds are of same size and the strengths of magnetic
field within different clouds are of same order. As a result one
may write:

\begin{equation}
p_{av}(\%) \sim constant +
\frac{1}{\sqrt{T_g}}(\frac{1}{T_d}-\frac{1}{T_g})
\end{equation}

 However, the
classical Davis \& Greenstein Mechanism has undergone many
modifications and various other grain aligning mechanisms are now
being used to explain the background star polarization (Cugnon
1985; Lazarian 1997; Lazarian et al. 1997). In the present work
restricting oneself to the simplest classical model of Davis \&
Greenstein, one should get the polarization observed in a cloud to
be related to the dust and gas temperature ($T_d$ and $T_g$
respectively) by the Eqn.(6).

In the present analysis  average polarization values for nine
(=8+1) clouds  are available. One can study the dependence of
these polarization values on the  dust and gas temperatures ($T_d$
and $T_g$). The average polarization values as observed for
different clouds $p_{av}$  are listed in Table 1. The values of
$T_d$, $T_g$ , as obtained from Clemens et al. (1991) are also
listed in Table 1. These authors used deep IRAS image analysis and
$^{12}CO$ spectroscopy to calculate dust and gas temperatures.
They calculated fluxes at 12, 25, 60 and 100 $\mu m$ bands and the
spectrum was not found to fit a single black body. This resulted
different temperatures for different band pairs, which was
explained  as the IR emissions coming from many different dust
populations  each at somewhat different temperatures. This
according to the authors may be expected, as the shape of the
interstellar extinction curve justifies a range of dust grains
sizes as shown by Mathis et al. (1977). One can see from Eqn (6)
that $p_{av}$ has a linear dependence on  $1/(T_d)$. Thus if
different dust populations give rise to different values of
polarization, one may consider an effective value of polarization,
calculated based on the harmonic mean of three dust temperatures.
The values of harmonic mean of the three dust temperatures
$T(12/25)$, $T(25/60)$ $T(60/100)$ as listed by Clemens et al.
(1991) are calculated and these mean values are actually listed
as $T_d$ in Table 1.

The same authors from their CO spectroscopy have also determined
the radiation temperature ($T_R$) for all the CB clouds, which
have been converted into gas  kinetic temperature in some cases
following the procedure as laid out by Dickman (1978). By
following the same procedure the gas kinetic temperatures ( $T_g$
as in  Eqn. (6)) for the present sample of clouds were calculated
and the values are listed in Table 1. From the values of $T_d$ and
$T_g$ the value of the expression
$\frac{1}{\sqrt{T_g}}(\frac{1}{T_d}-\frac{1}{T_g})$ is also
evaluated and denoted  by $T1$.

 In order to calculate the average of polarization
and position angle  values, one can estimate the weighted mean,
where the weights  are inverse of the square of errors $e_p$ and
$e_{\theta}$ respectively. However one may note that, stars which
are background to the cloud are fainter (due to extinction) and
thereby will have higher values of $e_p$. On the other hand the
foreground stars will have lower values of $e_p$. Therefore if one
weighs the data with inverse of $e_p$ or $e_\theta$, then attempt
to model fit the cloud polarization will give more emphasis on the
foreground stars, rather than the background ones. These will
clearly defeat our purpose of analysing the polarizing properties
of the cloud. With this justification only simple un-weighted
averages of $p$ and $\theta$ values are considered for this
analysis.

In Fig. 3, the average polarization ($p_{av}$) values are plotted
against the $T1$ values. For CB62 and CB246, $T1$ values were not
calculated as data was not available from Clemens et al. (1991).
As can be clearly seen from Fig. 3, the plot does not suggest any
relation between $p_{av}$ and $T1$ as is expected from Eqn. (6).
However, with the present condition that $T_g < T_d $, one expects
the long axis of grains to be aligned parallel to the magnetic
field. The negative  polarization values as obtained through the
present analysis (for the given set of gas and dust temperature
values) indicate such a geometry of alignment.

Lazarian et al. (1997)  in their work on the dark cloud L1755
tried to explain the polarimetric data in terms of grain alignment
mechanisms other than Davis \& Greenstein. Considering the grains
to be of super-paramagnetic material (with an justification from
Goodman \& Whittet 1995), the authors suggested that the degree of
Davis \& Greenstein alignment should depend on $(T_d / T_m)$ where
$T_m=(T_d + T_g)/2$ or the average of  of dust and gas
temperatures. Thus one should have

$$ p (\%) \sim (\frac{T_d}{(T_g + T_d)})  $$

Based on above  a plot of $p_{av}$ versus $\frac{T_d}{(T_g +
T_d)}$ ( denoted by  $T_2$) is made as in Fig. 4. This plot also
does not show any systematic dependence of polarization on
temperature (in terms of the parameter $T_2$). However, if one
excludes the data corresponding to CB4, it appears that a straight
line ($p \sim T_2$) may be fitted. At least  compared to Fig. 3,
the present plot in Fig. 4 (excluding CB4) shows  some indications
for a dependence of $p$ on $T_2$, as expected. Also there can be
reasons for the exclusion of data corresponding to CB4, where
polarization values were obtained in V  filter rather than in
white light as in all other cases. The polarization in white light
is always lower than what is observed through band pass filters
(as polarization at different wavelengths combine vectorially to
give lower net average polarization). At this stage it may be also
noted that, there are mechanisms other than Davis- Greenstein one,
which are now being used by several authors to explain
polarization caused by aligned grains. These include Purcell
alignment, alignment by radiation torque, mechanical alignment of
suprathermally rotating grains (for  a detailed review please see
Lazarian et al 1997).

We may further note that, about half of the cloud sample here have
active ongoing star formation present within these clouds.
Embedded stars warm their local dust and thereby dominate the
luminosity  budget of their clouds. Thus the IRAS fluxes measured
in such clouds reflect the dust heated by the embedded stars and
not the quiescent cloud material. Also one may note, star
formation affects the gas in a cloud in many important ways. It
can similarly heat the gas traced by $CO$ and it will effect the
line width.

Thus there may be several reasons, due to which the above observed
average polarization $p_{av}$ fails to show any systematic
dependence on dust and grain temperatures.

\subsection{The dependence of polarization on the turbulence in the
cloud} In the present analysis it is observed that the average
polarization $p_{av}$ varies substantially from cloud to cloud
which have different physical conditions as listed by Clemens et
al (1991).  These authors also listed $^{12}CO $ line width (in
terms of $\Delta V~~ km ~~sec ^{-1}$) values, which are assumed to
be  good indicators of turbulence within the cloud (listed in
Table 1). Based on this the authors have also classified the
clouds into three groups : A ( $T < 8.5^{\circ} $ K and $\Delta V
< 2.5$ km sec $^{-1}$ ), B $(T > 8.5^{\circ} K)$ and C ($T <
8.5^{\circ} K$ and $\Delta V
> 2.5$ km sec $^{-1}$).

One may expect the turbulence to have its influence on the grain
alignment.  However, as pointed out by the anonymous referee of
this paper, turbulence does not disturb the grain alignment on the
scale one is dealing with here. As a result of fast Larmor
precession  the grain follows the local direction of magnetic
field if the time scale for change of magnetic field direction is
longer than the time of Larmor precession. Turbulence gives rise
to variations in the local magnetic field direction along the line
of sight and the alignment takes place in respect to the local
magnetic field.  In the line of sight there may be several
independent directions of alignment causing a net depolarization
and resulting a low value of observed polarization. In this
scenario an empirical relation of the type $ p=a~~exp(- \Delta V.
b)$ may be used to analyse the present situation. Here as the
turbulence becomes too high,  one should get a zero value for
polarization, even if other parameters (contained in $a$) are
favourable to produce high polarization. On the other hand if no
turbulence is present, one can not get 100 \% polarization as the
other parameters will decide the minimum observable polarization
(decided by the value of $a$). The above equation $ p=a~~ exp(-
\Delta V. b)$ can be linearised and by the method of least-square
one may fit the following curve to data

\begin{equation}
ln(p)= 1.0831- 0.2424 ~~\Delta V
\end{equation}

or

$p=2.95  exp(-0.24 ~~\Delta V)$

\medskip
 The above relation suggests a maximum value $2.95 \%$ for
background star polarization. Fig 5. shows a plot of $ln(p)$
versus $\Delta V$ along with the above line of best fit (Eqn. (7))
for all the clouds except  CB62 for which data is not available.
The data on CB4 is also included from Kane et al. 1995 in this
plot. In Fig 5. a clear trend is observed where the average
polarization decreases with increase in turbulence ${\Delta V}$.
This can be clearly explained, as one knows the turbulence present
in the cloud can be held responsible for  lowering of polarization
values. However, in Fig. 3 and 4 where $p_{av}$ has been plotted
across some meaningful function of $T$, no such clear relation
exists. But one may note that, if $p_{av}$ has a stronger
dependence on ${\Delta V}$ as compared to $T$, then one can not
explore the relation between $p_{av}$ and functions of $T$ from
Fig. 3 and 4. To analyse this situation little further, one may
study the relation by plotting data points $ln (p_{av}/T_2)$
against $\Delta V$, which will remove the effect of temperature
from the observed polarization.  A new straight line
$ln(p_{av}/T_2)=1.1848 -0.2457*\Delta V $ may be fitted on this
new set of data with no further reduction in fitting error. Also
one may note in the present case, corresponding to CB246  there
will be no data point. Therefore, this situation is not considered
any further.

\begin{figure}
   \centering
   \includegraphics[width=8cm]{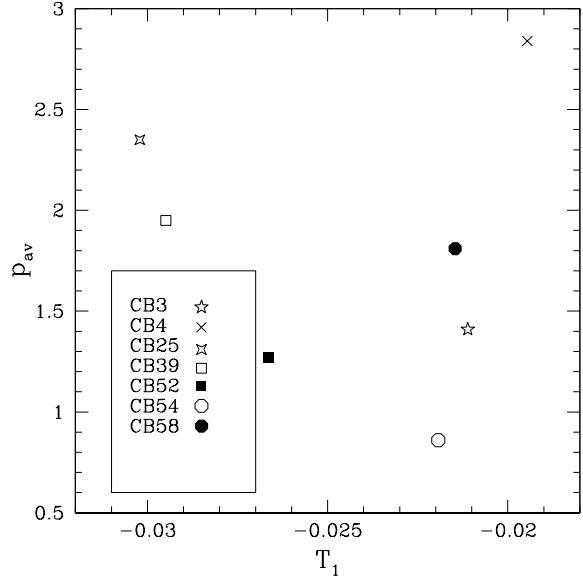}
      \caption{The average of observed polarization  ($p_{av}$) versus
      $T1 (= \frac{1}{\sqrt{T_g}}(\frac{1}{T_d}-\frac{1}{T_g}))$
      for various  clouds. }
         \label{FigVibStab}
   \end{figure}

\begin{figure}
   \centering
   \includegraphics[width=8cm]{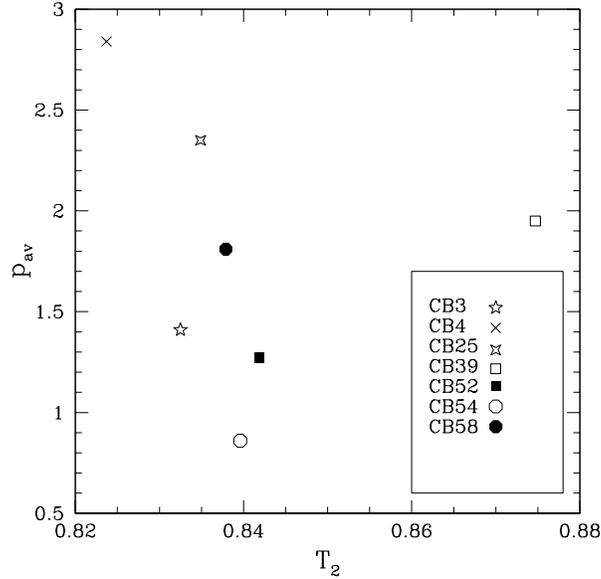}
      \caption{The average of observed polarization ($p_{av}$) versus
      $T2 (=\frac{T_d}{(T_g + T_d)})$
      for various  clouds. }
         \label{FigVibStab}
   \end{figure}

\begin{figure}
   \centering
   \includegraphics[width=8cm]{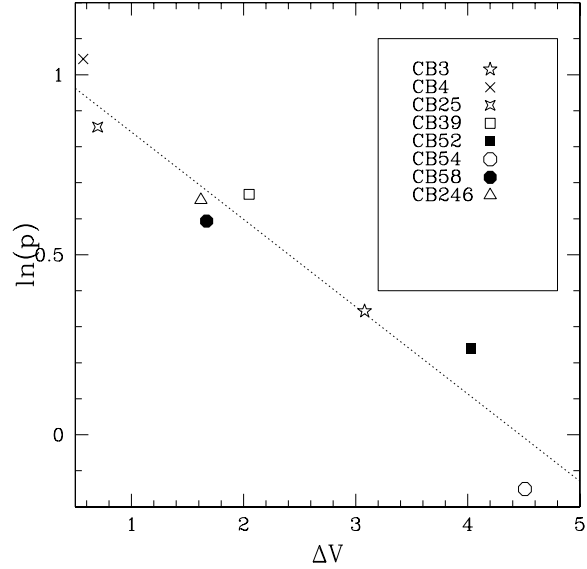}
      \caption{The log of average of observed
polarization  $ln (p_{av})$ are plotted against  the turbulence
$\Delta V$ for various clouds.  The line of best fit $ln (p)=
1.0831- 0.2424 \Delta V$ is shown along with.} \label{FigVibStab}
   \end{figure}

\begin{figure}
   \centering
   \includegraphics[width=8cm]{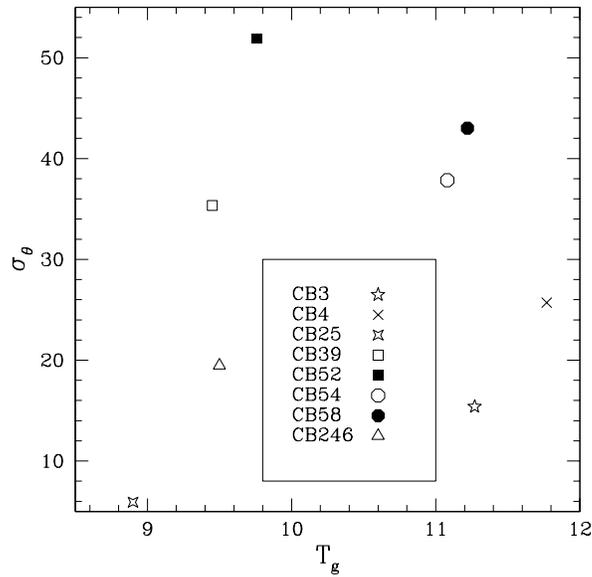}
      \caption{The dispersion in the direction of polarization
vectors ($\sigma _{\theta}$) are plotted against gas temperatures
($T_g$) for different clouds}
         \label{FigVibStab}
   \end{figure}

\begin{figure}
   \centering
   \includegraphics[width=8cm]{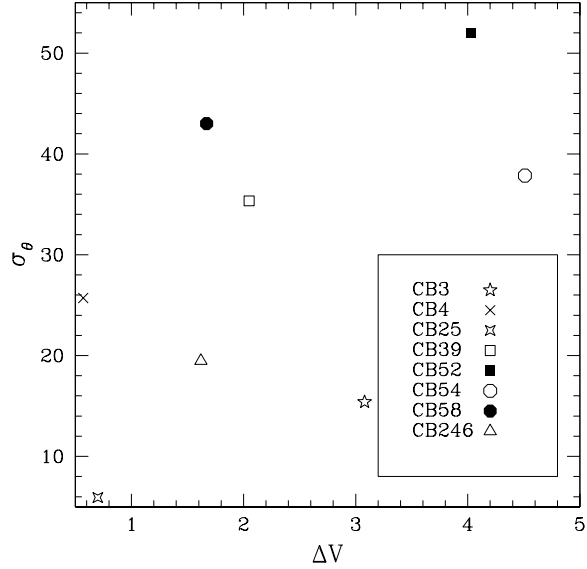}
      \caption{The dispersion in the direction of polarization vectors ($\sigma _{\theta}$)
      are plotted against amount of turbulence ($\Delta V$) for different  clouds}
         \label{FigVibStab}
   \end{figure}

\begin{figure}
\centering
\includegraphics[width=8cm]{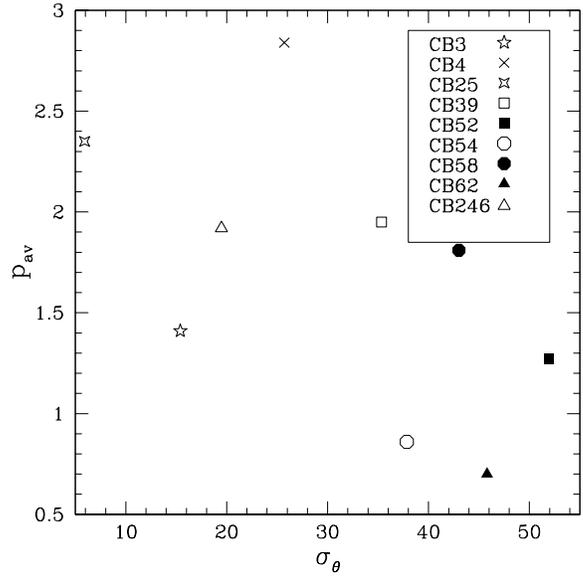}
\caption{The average of observed polarization ($p_{av}$) are
plotted against variance ($\sigma _{\theta}$) in the direction of
polarization vector } \label{FigVibStab}
\end{figure}

\begin{figure}
\centering
\includegraphics[width=8cm]{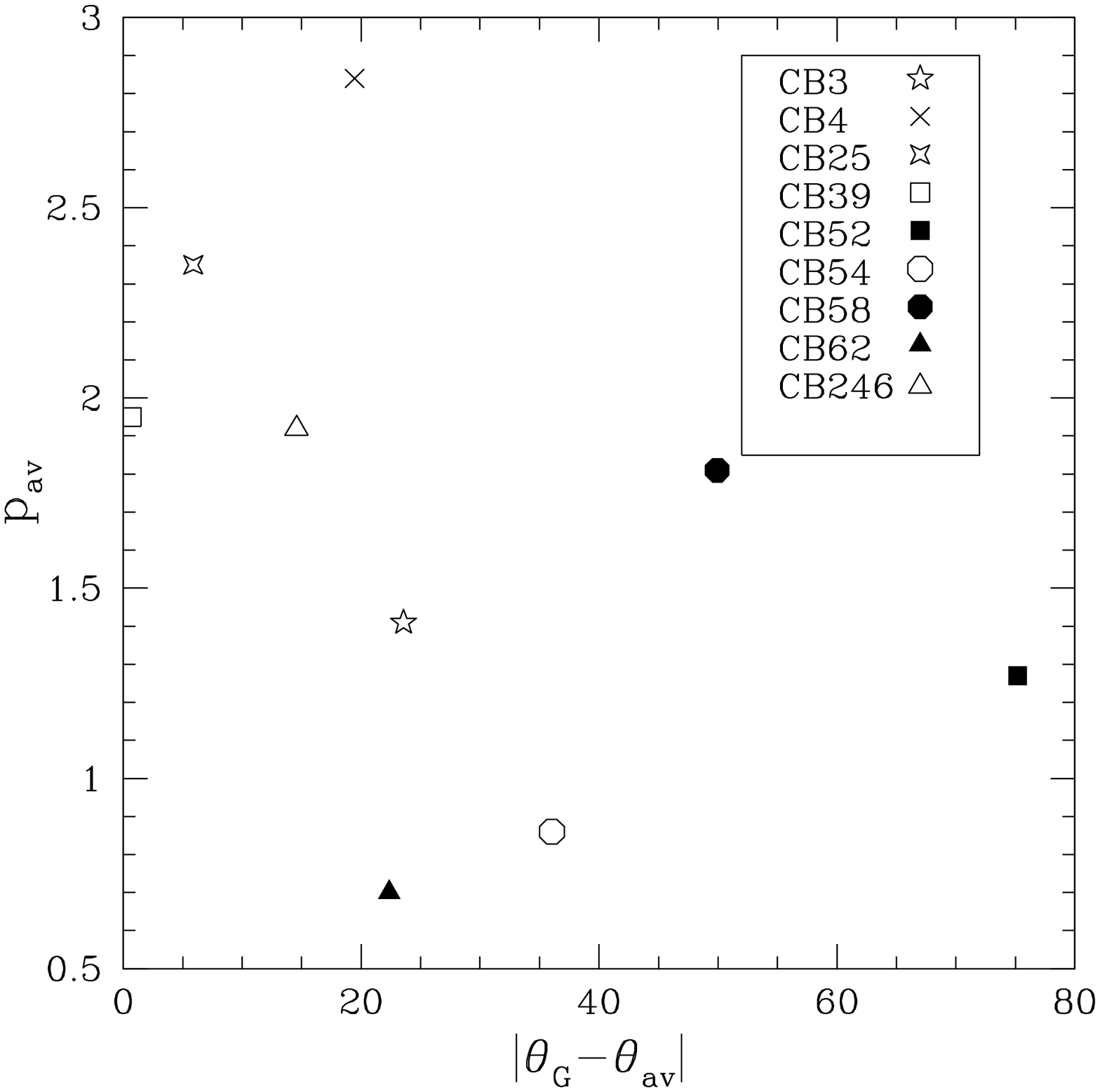}
\caption{The average of observed polarization ($p_{av}$) are
plotted against $|\theta _G - \theta_{av}|$} \label{FigVibStab}
\end{figure}

The dispersion ($\sigma _{\theta}$) in the direction of
polarization vector should be indicative of the non-uniformity in
the direction of the magnetic field over different parts of the
cloud. Now since this non-uniformity is observed in a plane
perpendicular to the line of sight, one can also expect
non-uniformity of same type to be present along the line of sight.
However, as pointed out by the anonymous referee, one notes  that
the non-uniformity in a plane perpendicular to the line of sight
is expected to be smaller than the  3-D non-uniformity along the
line of sight. The projection diminishes the plane-of-sky
non-uniformity. A random variation in the direction of magnetic
field should reduce the value of observed polarization, averaged
over the line of sight. The dust particles in a cloud can in
general be expected to be aligned by uniform interstellar magnetic
field. In that case one should have lower value of variance in the
direction of polarization. Also the average value of polarization
in the cloud should be close to the interstellar polarization
value in that part of the sky. But in situations where there exist
additional aligning mechanisms (operating within the cloud like
molecular outflow, etc.) one should get a higher dispersion
(variance) in the direction of polarization . This will happen
because the interstellar magnetic field will act in vectorial
combination with those additional aligning forces in the cloud. A
higher value of dispersion in $\theta $ is also possible if there
are irregularities (complexities) in the magnetic field structure,
caused due to mechanisms intrinsic to the cloud.

Thus a high value of $\sigma _{\theta}$ in some clouds, should
indicate that there are irregularities in the structure of
aligning forces. Also under such cases in these clouds, one should
get lower values of average polarization. In Fig 8.  $p_{av}$ has
been plotted against $\sigma _{\theta}$, where one can see
tendencies for a decrease in $p_{av}$ with the increase in $\sigma
_{\theta}$. In fact the data seem to be distributed into two
clusters, where two different straight lines may be fitted. The
data points corresponding to the clouds CB4, CB39, CB52 and  CB58
may be fitted into a separate straight line as compared to other
clouds. These two groups of data points however, do not
systematically fall in any of the groups A, B, C as discussed
earlier ( Cf. Table 1). One may try to explore the physical
properties of the two sets of clouds which are responsible for
this observed behaviour. This is planned for future work.

If grains are aligned by galactic magnetic field, one should find
average $p_{av}$ in a cloud to be higher where the difference
between the galactic plane direction  $\theta_G$ and $\theta
_{av}$ ie $|\theta_G -\theta_{av}|$ is lower. To establish this
idea in Fig. (9), $p_{av}$ has been plotted across $|\theta_G -
\theta _{av}|$ for different clouds. Primarily it appears to be a
scatter plot, but one can find tendencies  for the increase in
$p_{av}$ with the decrease in $|\theta_G - \theta _{av}|$ as
expected. Here also one can notice that, the data points
corresponding to the clouds CB4, CB52 and CB58 can be fitted into
a separate straight as compared to the other clouds which fit in a
second straight line. One may note that CB52 and CB58 are two such
clouds which showed bimodal distributions for $\theta$ (Cf. Fig 2.
and Section 2), as compared to others.

From the above discussions and discussions in Sect 3.1 and 3.2, it
appears that the presence of turbulence lowers the polarization
observed in a cloud and thus can be accepted as one of the factors
responsible for the non-uniformity in magnetic field, disturbing
the grain alignment. As a result it can be concluded that
polarization observed for stars background to a given cloud, is
not independent of the physical properties of that cloud. The work
by Goodman et al. (1995) expressed concern that, it seems the
polarization observed for stars background to a cloud is
independent of the cloud and not produced within the cloud.

\begin{center} { Table 1. For various CB clouds, the number
of stars, average polarization, average position angle,
dispersion, dust and gas temperatures, turbulence , difference
$|\theta_G- \theta_{av}|$ and  cloud groups are shown }
\begin{tabular}{c c c c c c c c c c}

\\
\hline
 \textit{Name of}& \textit{No. of}& $p_{av}$& $\theta_{av}$
 & $\sigma_{\theta}$
  & $T_d (^0 K)$ & $T_g (^0 K)$& $\Delta V (km s^{-1}) $& $|\theta_G - \theta_{av}|$&\textit{Cloud Group}\\
  \textit{the cloud}& \textit{stars}& & & & & & &\\
\hline
CB3   & 31 & 1.41  & 65.43 & 15.40 & 56   &11.27   &3.08 &23.57 &C\\
CB25  & 21 & 2.35  &150.89 & 5.91  & 45   &8.90    &0.70 &5.89  &A\\
CB39  & 21 & 1.95  &150.27 & 35.35 & 66   &9.45    &2.05 &0.73  &A \\
CB52  & 16 & 1.27  & 77.81 & 51.95 & 52   &9.76    &4.03 &75.19 &C \\
CB54  & 48 & 0.86  &115.96 & 37.85 & 58    &11.08   &4.51 &36.04 &C \\
CB58  & 29 & 1.81  &101.06 & 43.01 & 58   &11.22   &1.67 &49.94 &A\\
CB62  & 13 & 0.70  & 67.64 & 45.80 & 51   & --     &--   &22.36  &?\\
CB246 & 14 & 1.92  & 67.43 & 19.48 & --    &9.50    &1.62 &14.57  &A\\
CB4   & 80 & 2.84  & 70.55 & 25.71 & 55   &11.77   &0.57 &19.45  &A\\

\hline
\end{tabular}
\end{center}


\section {The spatial distribution of the polarization and position angle values }

\subsection{ A simple model for the polarization introduced by the
cloud}

The clouds which have been observed by Sen et al (2000) are nearby
with distances less than 600 pc (Clemens \& Barvainis (1988)). As
a result the polarization observed for these background stars can
be assumed to consist of only two components:

1) polarization introduced by the interstellar ( IS ) medium
background to the cloud. For the foreground stars as the cloud is
nearby, one can neglect the polarization caused due to the IS
medium between the cloud and the observer.

2) polarization introduced by the cloud itself, which is believed
to be optically thicker.

It is assumed  that the polarization properties of the IS medium
can be approximated by  a simple dichroic sheet polarizer. The
transmission properties of a simple dichroic sheet polarizer
(aligned in an arbitrary direction by angle $\phi$ with respect to
some reference direction ) can be mathematically represented by
the following Mueller Matrix (Kliger et al 1990; Shurcliff 1962)

\begin{equation}
 A(\phi)= {1 \over 2}\pmatrix{ (k_1+k_2)    &(k_1-k_2)c_2
&(k_1-k_2)s_2 &0 \cr (k_1-k_2)c_2 &(k_1+k_2)c_2^2+2s_2^2k
&(\sqrt{k_1}-\sqrt{k_2})^2c_2s_2        &0 \cr (k_1-k_2)s_2
&(\sqrt{k_1}-\sqrt{k_2})^2c_2s_2 &(k_1+k_2)s_2^2+2c_2^2k &0 \cr 0
& 0 & 0 &2k \cr}
\end{equation}

where $c_2=cos(2\phi)$ ; $s_2=sin(2\phi)$; $k=\sqrt{k_1 k_2}$ and
$ k_1$ and $k_2$ are the transmission coefficients for light when
the electric vector is parallel and perpendicular to the optic
axis of the polarizer.

It is assumed the light from the background star is initially
unpolarized and so it can be represented by the Stokes coloumn
matrix (Shurcliff 1962; Stoke 1852) $S= \{I~~ 0~~ 0~~ 0 \}$. If
one assumes the IS medium to be a dichroic polariser with optic
axis making an angle 0 (zero) with the reference  direction then
one can represent its polarizing properties by the Mueller matrix
$A(0)$. Similarly the cloud can be represented by the matrix
$A^{\prime} (\phi)$ with transmission coefficients $k_1 ^{\prime}$
and $k_2 ^{\prime}$. Now let the light reaching the observer after
it comes out of the cloud, be represented by the Stokes coloumn
matrix $S'= \{ I' ~~ Q' ~~ U' ~~ V' \} $. Therefore one may write

\begin{equation}
\lbrack S' \rbrack= \lbrack A'(\phi) \rbrack\lbrack A(0)
\rbrack\lbrack S \rbrack
\end{equation}

After the appropriate matrix multiplication one gets
\begin{equation}
\pmatrix{I'\cr Q'\cr U'\cr V'\cr}={1\over 4}
\pmatrix{I(k_1+k_2)(k'_1+k'_2) + I(k_1-k_2)(k'_1-k'_2)c_2 \cr
        I(k_1+k_2)(k'_1-k'_2)c_2 +I(k_1-k_2)((k'_1+k'_2)c_2^2+ 2s_2^2 k')\cr
                I(k_1+k_2)(k'_1-k'_2)s_2 +I(k_1-k_2)(\sqrt{k'_1}-\sqrt{k'_2})^2c_2s_2\cr
                        0\cr}
\end{equation}
where $k'=\sqrt {k'_1 k'_2}$. After simplification the above
Eqn.(10) reduces to

\begin{equation}
\pmatrix{I'\cr Q'\cr U'\cr V'\cr}={Ik_1k'_1(1+f)\over 4}
\pmatrix{(1+f')+p(1-f')c_2 \cr
        (1-f')c_2 +p((1+f')c_2^2+ 2s_2^2\sqrt{f'})\cr
                (1-f')s_2 +p(1-\sqrt{f'})^2c_2s_2\cr
                        0\cr}
\end{equation}

where $f=k_2/k_1$  and $f'=k'_2/k'_1$.

Since the interstellar polarization $p=(k_1-k_2)/(k_1+k_2)$,  one
may also write
\begin{equation}
f=(1-p)/(1+p)
\end{equation}

\subsection{A model for the transmission coefficients of the cloud:}

In general one may assume the cloud is spherical in shape, with
radius $R_0$. Now as shown in Fig 10, the background star is seen
through the cloud, at a radial  distance $r$ from the center of
the cloud. Therefore, the starlight passes a distance $ 2h$
through the cloud, where $h \sim \sqrt{R_0^2-r^2}$. One may assume
that the starlight passes through '$2h$' number of layers through
the cloud and the polarizing effect of each layer is equivalent to
$c$ (some arbitrary constant) times that of the IS medium, where
the later has transmission coefficients $k_1$ and $k_2$. This also
amounts to the assumption that the composition (characterised by
$k_1$ and $k_2$) of the dusts in the cloud and the IS medium are
the same. However, within the cloud the grains may have higher
number density or may be better aligned, introducing a higher
amount of polarization in the light from background stars. Grains
may be also aligned in a direction different from that in IS
medium. This is the simplest possible model which is considered
for the present analysis. Now since there are '$2h$' number of
such layers, the equivalent transmission coefficients for the
cloud would be $(k_1)^{2hc}$ and $(k_2)^{2hc}$. In other words one
writes $k'_1=(k_1)^{2hc}$ and $k'_2=(k_2)^{2hc}$. The estimated
(or expected) value of polarization $pe$ present in the light
coming out of the cloud can be  expressed as
$pe={{\sqrt{Q'^2+U'^2}}\over {I'}}$ .

Thus with the help of Eqn. (11) one writes :

\begin{equation}
pe = {{\sqrt{((1-f')c_2+p((1+f')c_2^2+2s_2^2\sqrt{f'}))^2
+((1-f')s_2+p(1-\sqrt{f'})^2c_2s_2)^2}}\over {(1+f')+p(1-f')c_2}}
\end{equation}

At this stage one can consider two special cases when $\phi = 0$
\& $90 $ degrees and these cases are represented by the following
two equations:
$$pe(\phi=0)={{(1-f')+p(1+f')} \over {(1+f')+p(1-f')}}$$

$$pe(\phi=90)={{-(1-f')+p(1+f')} \over {(1+f')-p(1-f')}}$$

However in general when $p\ll 1$, one can use the approximation
$f'=f^{2h}=((1-p)/(1+p))^{2h}\simeq (1-2hp)/(1+2hp)$ and from Eqn.
(11) one can write

$pe \simeq$
$${{\sqrt{(2hpc_2+p(c_2^2+s_2^2(1-hp)(1+2hp)/(1+hp)))^2+
(2hps_2+p(1-(1-hp)(1+2hp)/(1+hp))c_2s_2)^2}}\over {1+2hp^2c_2}}
$$

\subsection{Fitting the observed polarization for radial distance from cloud centre}

For all the eight clouds observed , the radial distances $r$ (in
arc sec) have been estimated from the co-ordinates (RA and DEC) of
each star available in Sen et al. (2000). The polarizations
observed for such field stars in white light are plotted against
$r$, in Figs. (11) and (12). In some cases  there is a trend
(example CB25, CB39), where as one moves away from the cloud
center the polarization decreases and then attains a minimum value
somewhere between 150-250 arc sec. After that as one  moves toward
the periphery of the cloud, the polarization value increases and
reaches the IS polarization value as one finally moves out of the
cloud.

In order to find some estimate for the interstellar polarization
for the nearby region of the cloud, one can take a vectorial
average of all the polarization values (taking into account the
associated position angles in the measurements) that have been
observed for the outer most part of the cloud and assume that
value to be representative of the IS polarization value (denoted
by $p$, cf. Table 2).  The same Table also lists the distances
corresponding to the outer most star, which is also assumed to be
roughly the dimension  $R_{0}$ of the cloud.

The values of $r$ and $h$  assumed by us,  in principle should be
proportional to the actual distances within the cloud. The
quantity $c$ in the expression $h=c\sqrt{R_o^2-r^2}$ (as was
introduced earlier in Section 4.2) can act as a proportionality
constant to  normalise  such  distances. The value of $c$ will be
optimised later during model fitting the data.

After finding out $r$ (and hence $h$) for each field star, one
estimates the polarization $pe$, using relation (13) and  minimize
the value of quantity $\chi^2= \sum{{(pe-po)^2}}$ with $c$ and
$\phi$ as fitting parameters (where $po$ is the observed
polarization). While minimising the value of $\chi^2$, the data
was not weighted with $1/(e_p)^2$, the justifications of which
have been already discussed in Section 3.1. In Table 2. one finds
the optimized values of $c$ and $\phi$, along with the
corresponding minimised value of $\chi^2$. In this connection it
may be mentioned that, while minimising the values of $\chi^2$, it
was noticed that only at a unique set of values for ($c$ and
$\phi$), the minimum value for $\chi^2$ can be obtained.

Fig. 11  shows a  plot of the polarization values (po) observed
for all the field stars in the clouds CB3, CB25 , CB39 and CB52,
along with a curve representing the polarization  ($pe$) values as
estimated using Eqn. (13) and after best values of $c$, $\phi$
have been obtained by $\chi^2$ minimisation technique. Fig 12
represents a similar plot for CB54, CB58 , CB62 and CB246.

From these two set of figures one finds that  for some clouds like
CB3, CB25 and CB39 the polarization value is relatively high at
near the center region and then reaches a minimum at a distance
varying between 150-250 arc sec. After this it increases again as
one moves out of the cloud  and reaches the region of IS medium.
At least for the clouds CB25 and CB39 this feature is more clearly
seen. This happens when the relative angle between the magnetic
field in cloud (related to the direction of optic axis) and that
of IS medium, $\phi$ is close to $90 ^{o}$.

For the other clouds CB52, CB54, CB58, CB62 and CB246 one can see
this ideal  model does not fit to the data. Thus once can rule out
the possibility that these clouds can be represented by a simple
sphere containing an uniformly directed magnetic field,
responsible for the alignment of grains. However, even an ideal
cloud of above type may not fit to the observed data due to any
(or all) of the following reasons :

1) There may be always some stars, which are foreground to the
cloud. They will however, exhibit very little polarization, unless
they are intrinsically polarized.

2) Some of the background stars, even may show high difference in
polarization from the model curve, if the stars themselves are
intrinsically polarized.

3) All the stars, background to the cloud are not placed at same
distances behind the cloud. As a result they are passing through
different distances through the IS medium and will have different
values of IS polarization in them.

4)If the shape of the cloud is different from an ideal sphere.

As at present it is not possible to distinguish the background
stars from the foreground ones, the 'goodness of fit' can not be
improved any further.

The goodness of our fit could  have been also improved, if one
could have measured polarization values sharply at a particular
wavelength (say through narrow band filters), rather than white
light. This is because the polarization produced by passage
through dichroic polarizer, as discussed above, has strong
wavelength dependence and stars are of various spectral types.

However, based on the present analysis one can not claim that a
uniformly directed magnetic field (for that matter any aligning
force) exists throughout the entire cloud which is assumed to be
spherical, with exceptions like in CB3, CB25 and CB39. In clouds
CB25 and CB39 ( and to some extent for CB3)  the magnetic field
appears to be quite uniform.

Further based on the present analysis one can also show that, the
curve relating $pe$ with $r$ can assume different shapes,
according to different values of $\phi$. So it is not always
necessary that, as one moves towards the centre of the cloud, the
polarization should also increase. Goodman et al. (1995), had
questioned  the validity of background star polarimetry as a tool
to study the cloud properties. The main concern expressed by the
authors was that as one moves towards the interior (center) of the
cloud the total extinction ($A_v$) increases, but the polarization
does not increase as expected.  However, with the help of present
analysis one can show that, the observed polarization depends
largely on the geometry of the magnetic field (as aligning force)
within the cloud and as a result it does not always increase with
$A_v$.

\begin{figure}
   \centering
   \includegraphics[width=8cm]{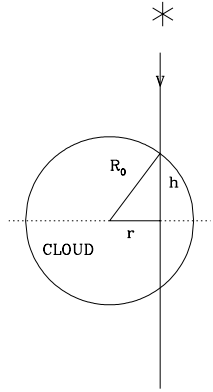}
      \caption{A model
for cloud with the light from background star passing through it}
         \label{FigVibStab}
   \end{figure}

\begin{figure}
   \centering
   \includegraphics[width=8cm]{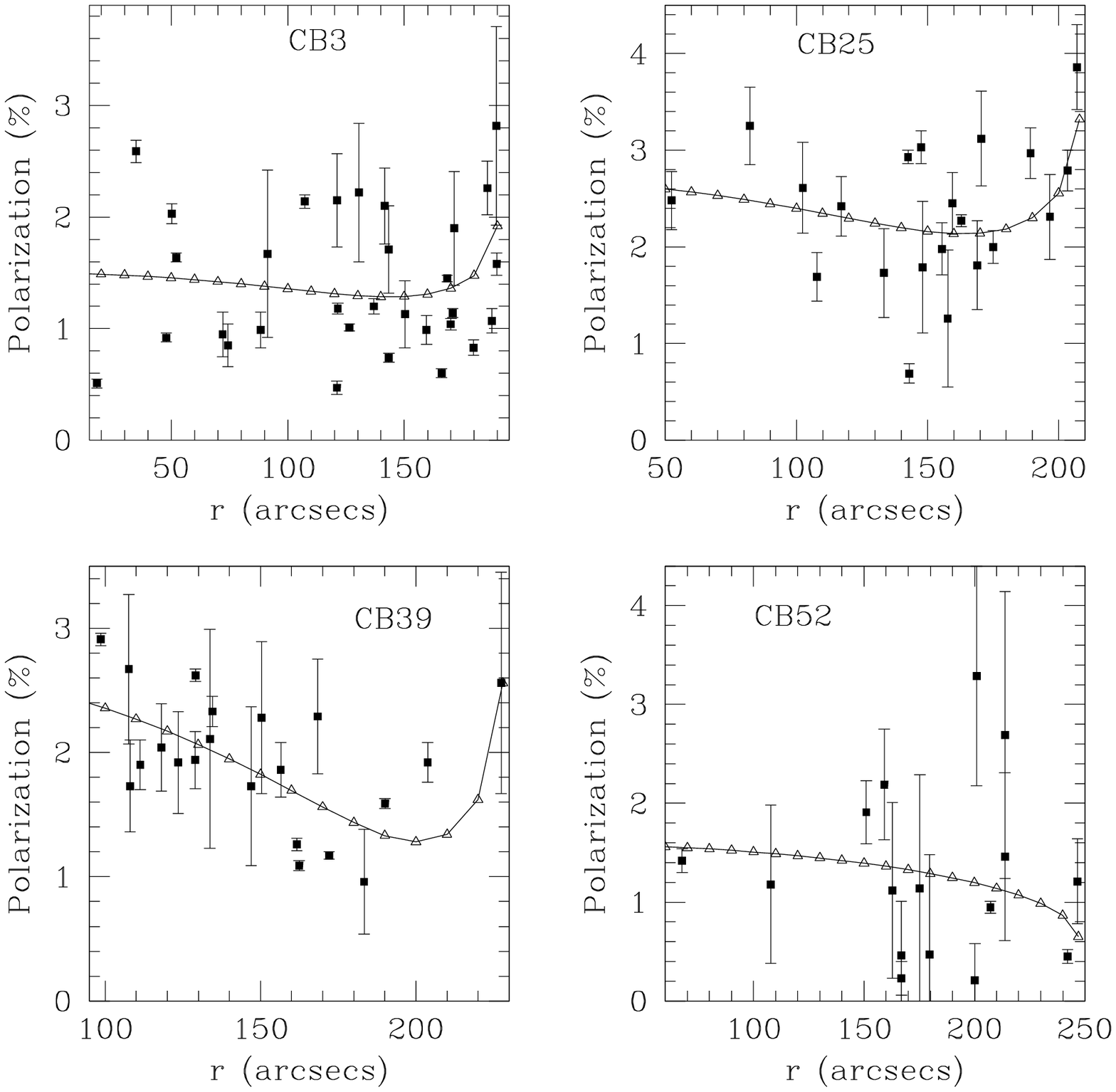}
      \caption{Observed
Polarization versus radial distance plot for the clouds CB3, CB25,
Cb39 and CB52. The curves joining the $\triangle$, represent our
proposed model.}
 \label{FigVibStab}
\end{figure}

\begin{figure}
   \centering
   \includegraphics[width=8cm]{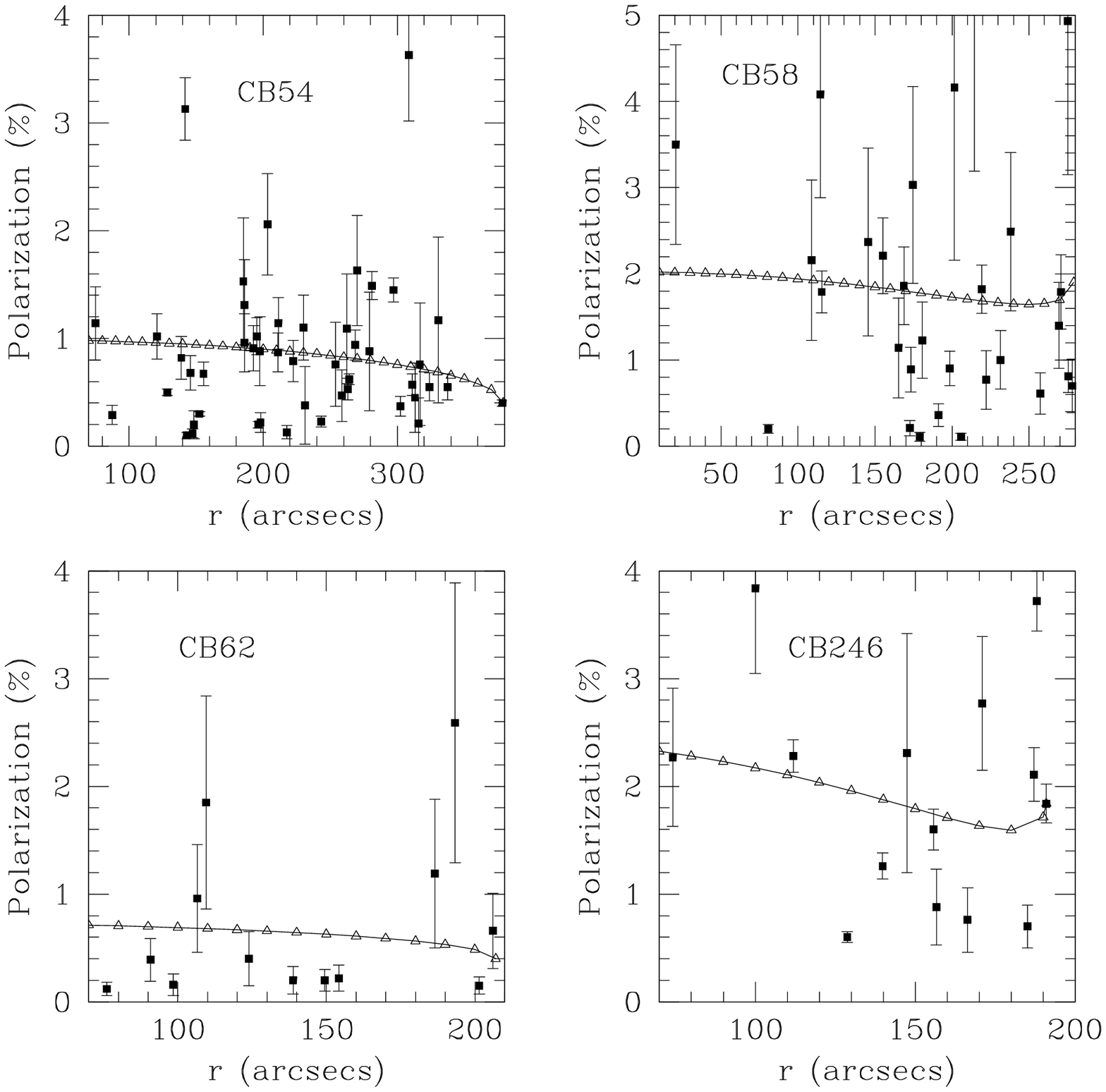}
      \caption{Observed
Polarization versus radial distance plot for the clouds CB3, CB25,
Cb39 and CB52. The curves joining the $\triangle$, represent our
proposed model.} \label{FigVibStab}
   \end{figure}

\begin{center} {Table 2. The values of $R_0$ (arc sec), interstellar polarization $p$ (in
\%), $\phi$ (in degrees), $c$, $\chi ^2$ are shown}
\begin{tabular}{lccccc} \hline \hline
Cloud  &$R_0$&$p$& $\phi$& c& $\chi ^2$\\ \hline
CB3  &190  &1.92  &69  & 0.003 & 11  \\
CB25 &208  &3.32  &70  & 0.003 & 8   \\
CB39 &228  &2.56  &75  & 0.004 & 4  \\
CB52 &247  &0.65  &10  & 0.003 & 12  \\
CB54 &379  &0.40  &10  & 0.002 & 23  \\
CB58 &279  &1.90  &60  & 0.002 & 62  \\
CB62 &207  &0.40  &0  & 0.002 &  8  \\
CB246&191  &1.84  &60  & 0.004 & 13 \\
\hline

\end{tabular}
\end{center}

\section{ Conclusions}

The polarization observed for stars background to eight clouds
(from Sen et al. 2000) and one from Kane et al. (1995) have been
analysed and some of the major conclusions are summarised  below:

1. A histogram plot showing  Rice corrected polarization values
against number of stars, shows bimodal distribution with two peaks
in polarization values, for some of the clouds. A possible
interpretation in terms of a mixture of polarization due to IS
medium and that due to cloud are discussed

2. A similar histogram plot with position angle ($\theta$) values,
also shows  some indications for bimodal distribution, which can
be explained in terms of the inhomogenities in magnetic field
geometry. However, the average direction of polarization vector
and that of the interstellar magnetic field seem to be the same.

3.The observed average polarization in a cloud does not appear to
be  related to the dust and gas temperatures as expected from
Davis \& Greenstein (1951) mechanism.

4.The observed average polarization ($p$) and turbulence ($\Delta
V$) present in the cloud, can be related  by a line of best fit
$ln(p)= 1.083- 0.2424 \Delta V$.

This finding bears importance as one can show that physical
conditions within the cloud can influence the polarization which
one observes for stars background to the cloud.

5. By assuming a given cloud  to be a simple dichroic sphere, one
can calculate the expected polarization values for stars  at
different projected distances from the cloud center. This model
can  explain to a reasonable extent the spatial distribution of
observed polarization in  CB25 and CB39 (and to some extent CB3).
But for other clouds the model fails.

However, based on this model one can explain why polarization
always does not increase with total extinction $A_v$ as one moves
towards the center of the cloud.

\section*{Acknowledgements}
We are thankful to Prof. Dan Clemens of Boston University, for
discussion (over email) on the calculation of gas kinetic
temperature in CO spectroscopy.

The authors thank JSPS, Japan and DST, India for providing funds
under their exchange programme which made this work possible.

We are also thankful to the anonymous referee of this paper for
valuable suggestions and views, which we believe have definitely
helped to improved the quality of this paper.

\end{document}